\documentclass[twocolumn,showpacs,preprintnumbers,amsmath,amssymb,floatfix]{revtex4}

\usepackage{graphicx}
\usepackage{dcolumn}
\usepackage{bm}

\begin{document}
\newcommand*{\bra}[1]{\ensuremath{
  \left\langle #1 \right\vert }}
\newcommand*{\ket}[1]{\ensuremath{
  \left\vert #1 \right\rangle }}
\newcommand*{\braket}[2]{\ensuremath{
  \left\langle #1 | #2 \right\rangle}}
\newcommand*{\E}[1]{\ensuremath{
  \left\langle #1 \right\rangle}}
\newcommand*{\partiald}[2][]{\ensuremath{
  \frac{\partial #1}{\partial #2} }}
\newcommand*{\totald}[2][]{\ensuremath{
  \frac{d#1}{d#2} }}
\newcommand*{\vect}[1]{\ensuremath{
  \mathbf{#1}}}
\newcommand{\displacement}[2][]{\ensuremath{
  \hat D^{#1} \negmedspace \left( \left\{ #2 \right\}\!\right)}}

\preprint{APS/123-QED}

\title{Entanglement generated between a single atom and a laser pulse}

\author{Andrew Silberfarb}
\email{drews@unm.edu}
\author{Ivan H. Deutsch}
 \email{ideutsch@info.phys.unm.edu}
\affiliation{Department of Physics and Astronomy, University of
New Mexico, Albuquerque, NM 87131}

\date{\today}

\begin{abstract}
We quantify the entanglement generated between an atom and a laser pulse in free space.  We find that the entanglement calculated using a simple closed-system Jaynes-Cummings Hamiltonian is in remarkable agreement with a full open-system calculation, even though the free-space geometry is far from the strong coupling regime of cavity QED. We explain this result using a simple model in which the atom couples weakly to the laser while coupling strongly to the vacuum.  Additionally we place an upper bound on the total entanglement between the atom and all paraxial modes using a quantum trajectories unravelling.  This upper bound provides a benchmark for atom-laser coupling.
\end{abstract}


\maketitle

\section{Introduction \label{S:intro}}

The electromagnetic field describing a laser pulse is intrinsically quantum mechanical and may become entangled with an atom when the two systems interact.  The consequences of this fact for experiments which employ lasers as control fields has been the topic of several recent papers \cite{vanEnk2001,Banacloche2002,Silberfarb2003}.  All agree that these quantum effects are negligible compared to errors induced by spontaneous emission out of the beam.  In particular, in our previous work \cite{Silberfarb2003}, we calculated the rate at which information is gained about a two-level atom when one measures a near resonance laser pulse that has interacted with it.  This measurement strength could be interpreted as the rate at which photons are spontaneously emitted into the modes defined by the laser beam -- a small fraction of the total emission rate.  Indeed, as commented by Itano \cite{Itano2003}, the quantum fluctuations in a laser beam are equivalent to those in the vacuum, and thus using the standard unitary Mollow transformation \cite{Mollow1975}, all quantum effects related to the atom-laser interaction in free space are derived from spontaneous emission in the transformed frame.

Though not an important source of error in quantum operations, entanglement is nonetheless generated between the atom and the laser beam.  Photons spontaneously emitted into the laser modes change the quantum statistics of those modes in a way that is correlated with the state of the atom.  This entanglement might prove useful as a resource in some quantum communication or quantum computation scheme\cite{Nielsen2000}.  Moreover, by determing how much entanglement one can generate in free space, we obtain a benchmark for comparison with other entanglement generation schemes.  For example, the entanglement generated in a cavity QED geometry \cite{Kimble1998} could be compared to the entanglement that would be generated in the absence of the resonator. 

Quantifying the entanglement generated between an atom and a laser in free space is complicated by the \textit{multimode} nature of the field in a traveling wave geometry. The Hilbert space describing the laser modes is the tensor product of many independent subsystems.  Additionally, the laser and atom do not form a closed system, as the other vacuum modes of the field interact with the atom.   Since spontaneous emission into these unoccupied modes dominates over quantum features within the laser modes, it is generally inconsistent to neglect their effects when calculating atom-laser entanglement.

To analyze this problem, we use the formalism developed in our previous paper \cite{Silberfarb2003}.  We review the salient features in the next section.  We then proceed to examine the entanglement generated between the laser pulse mode and the atom.  This result is compared to similar results obtained using a closed-system single mode picture.  We follow this with a calculation of an upper bound on the total entanglement generated between the atom and all modes propagating with the laser beam.  A brief summary of results is presented in the concluding section.

\section{Paraxial formalism \label{S:formalism}}
In many problems it is useful to have a local description of the quantized field.  For example, when the field interacts with a well localized atom or when an experiment uses a detector to continuously measure the field, a delocalized plane wave description is awkward.  In systems where a physical quantization volume is defined, such as in cavity QED \cite{Kimble1998}, a local set of quantized modes is naturally defined.  However, in free space no such intrinsic volume is present.  A quantization procedure which considers local temporal modes rather than delocalized spatial modes  \cite{Blow1990} has a limited regime of validity.  Instead, through an appropriate coarse-graining of the standard modal decomposition, we can achieve the desired local picture \cite{Silberfarb2003}.  A brief review of the pertinent points follows.

The procedure begins by separating the field modes into paraxial and nonparaxial modes.  In the canonical example  a laser beam impinges upon an atom, interacts with it, then illuminates a detector.   This optical system defines a set of modes traveling essentially in one direction with a small diffraction angle.  We quantize these paraxial modes by associating a Fock space to every member of a coarse-grained set of  wave packets of width $\Delta z$ and transverse area $A$.  These form an approximate complete set for the paraxial subspace if the dynamics on a time scale smaller than $\Delta t = c/\Delta z$ are negligible.  The coarse-graining length is arbitrarily chosen for convenience and generally should not appear in any experimentally relevant result.  It may be chosen to be the width of a trapped atom or to correspond to the detector bandwidth $\Delta \omega$, using $\Delta z = c/\Delta \omega$.  This latter definition is natural when considering the entanglement resource than can be detected.  We introduce coarse-grained spatially localized mode functions $\phi_i(z)$ and an associated set of annihilation operators  $\hat{a}_i$.  The Hamiltonian for the paraxial field and atom in the interaction picture and under the rotating wave approximation is,
\begin{eqnarray}
  \hat H_\text{AF}(t)=\hbar g\sum_i \Theta _i(t)\left( \hat a_i\hat \sigma _+
    + \hat a_i^\dag \hat \sigma _- \right)\label{E:Hamiltonian}\\
   \text{where } \hbar g=dE_{\text{vac}}
    =d\sqrt {\frac{2\pi \hbar \omega _0}{A\Delta z}}. \nonumber
\end{eqnarray}
Here $\Theta_i(t)$ are window functions selecting out the mode currently interacting with the atom
\begin{equation}
	\Theta _i(t)=\begin{cases}
    1,&(i-1)\Delta t<t\le i\Delta t \\
    0,& \text{otherwise}\label{E:Step}
  \end{cases}
\end{equation}
In this picture, each wave packet mode propagates past the atom, interacts with it according to a Jaynes-Cummings Hamiltonian \cite{Jaynes1963}, then flies away. These dynamics are intrinsically irreversible, in contrast to a cavity QED scenario in which excitaitons oscillate between the field and atom reversibly \cite{Kimble1998}.  The information contained in these modes can act to decohere the atom at a rate $\kappa \approx \Gamma \sigma_\text{eff} /A$, where $\Gamma$ is the total spontaneous emission rate and $\sigma_{\text{eff}}= 3 \pi/2 k_0^2$ is the effective scattering cross section on resonance.  The ``measurement strength", $\kappa$, represents the rate at which photons are spontaneously emitted into the paraxial modes. 

The nonparaxial modes which constitute the rest of the field are quantized using the standard free space quantization procedure, associating a Fock space to every plane wave.  In the following, these modes will be treated as a zero temperature reservoir into which the atom can decay.  The only difference from standard spontaneous emission will be that the rate of decay is slightly reduced by the removal of the paraxial modes.
\begin{equation}\label{E:gammap}
	\Gamma' = \Gamma \left(1-\frac{\Omega_p}{4 \pi}\right) \approx \Gamma - \kappa,
\end{equation}
with $\Omega_p$ the solid angle subtended by the paraxial beam.  Including the nonparaxial modes, the full master equation for the atom-paraxial field system becomes
\begin{equation}
	\totald[\hat{\rho}]{t} = \frac{-i}{\hbar} \left[ \hat{H}_\text{AF},\hat{\rho}\right] 
		-\frac{\Gamma'}{2} \left\{ \hat \sigma _+\hat \sigma _-,\hat{\rho}  \right\}
    +\Gamma' \,\hat \sigma _- \hat{\rho} \hat \sigma _+,
\end{equation}
where $\hat H_\text{AF}$ is the Hamiltonian that couples the paraxial field modes to the atom, given in Eq.\ (\ref{E:Hamiltonian}).  

In our canonical example the paraxial modes are taken to be in an initial coherent state, corresponding to the quantum state of a laser beam.  We perform a Mollow transformation  \cite{Mollow1975} on the above master equation to remove the initial coherent field amplitude of the laser and transfer it into the Hamiltonian.  In the ``Mollow picture'', the initial state of the field is the vacuum and the master equation becomes
\begin{equation}\label{E:Master}
	\totald[\hat{\rho}]{t} = \frac{-i}{\hbar} \left[\hat{H}_\text{coh} + \hat{H}_\text{AF},\hat{\rho}\right] 
		-\frac{\Gamma'}{2} \left\{ \hat \sigma _+\hat \sigma _-,\hat{\rho}  \right\}
    +\Gamma' \,\hat \sigma _- \hat{\rho} \hat \sigma _+,
\end{equation}
where $\hat H_\text{coh}$ is the coherent semi-classical Rabi-flopping Hamiltonian
\begin{equation}\label{E:Coherent}
	\hat H_\text{coh} = \hbar g \alpha \left(\hat \sigma_+ + \hat\sigma_-\right).
\end{equation}
The Mollow transformation is achieved by a local unitary operator acting on the field alone and so will not affect the atom-field entanglement.  Equation (\ref{E:Master}) thus serves as our of basic dynamical description for entanglement calculations.
 
\section{Quantifying Entanglement}
\begin{figure}
	\scalebox{.45}{\includegraphics{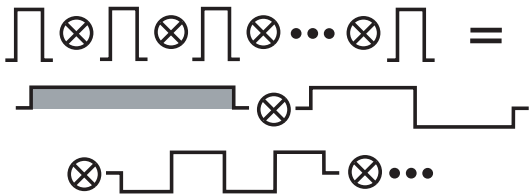}}
	\caption{Two different modal decompositions of the paraxial field.  The top line represents the local coarse-grained decomposition.  The second two lines depict the Fourier decomposition, which consists of many delocalized longitudinal modes. Only the symmetric mode (laser mode) is initially excited, starting in a  coherent state, and is shown here shaded.  The other nonsymmetric modes (two of which are depicted) are initially in the vacuum state. \label{F:Diagram2}}
\end{figure}

In reviewing the paraxial formalism in the last section we made a distinction between the paraxial and nonparaxial field modes, the paraxial modes being those modes which travel past the atom and impinge upon the detector.  We then traced out the nonparaxial modes to recover the master equation (\ref{E:Master}).  The remaining paraxial subsystem is still a very large dimensional Hilbert space, composed of  a tensor product of Fock spaces for each coarse-grained wavepacket in the pulse.  Such a large Hilbert space has many subsystem decompositions, any one of which could be of interest when calculating entanglement (Fig.\ \ref{F:Diagram2}).  In this section we consider two possible decompositions: (i) entanglement between the atom and the mode defined by the laser pulse,  (ii) entanglement between the atom and the entire paraxial subsystem.  The former is natural to consider when treating the laser pulse as a single mode, analogous to a Jaynes-Cummings cavity QED geometery \cite{Kimble1998}. The latter is of interest when relating the atom-laser entanglement to the measurement strength of the laser and to the error in quantum logic induced by a control pulse.  In addition, when calculated using this second division, we can find an upper bound on the atom-laser entanglement that can be used as resource for quantum information processing.  

To quantify entanglement, we use a monotone known as the ``tangle" which can be related to the square of the more familiar concurrence and also to the entanglement of formation \cite{Wootters1998,Osborne2002}.  For overall pure states where at least one of the systems has only two levels (is a qubit) the tangle is equal to the normalized linear entropy (or purity) of the marginal state of either subsystem \cite{Rungta2001},
\begin{equation}\label{E:PureTangle}
	T\left(\ket{\psi}\bra{\psi}\right) = 2 \left(1 - \text{Tr}_A \left[ \text{Tr}_B\left[\ket{\psi}\bra{\psi}\right]^2 \right] \right).
\end{equation}
This formula can be generalized to mixed states by taking the convex roof \cite{Uhlmann1998}
\begin{equation}\label{E:MixedTangle}
	T \left(\hat\rho \right) = \min_{\ket{\psi_i},p_i} \sum_i p_i T\left(\ket{\psi_i}\bra{\psi_i}\right)   \qquad  \hat{\rho} = \sum_i p_i \ket{\psi_i} \bra{\psi_i}.
\end{equation}
We choose to use the tangle instead of the full Von Neumman entropy due to its ease of calculation.
\section{Symmetric mode entanglement\label{S:Symmetric}}

\subsection{Calculation\label{S:SymCalc}}
We consider first the tangle between the atom an the ``symmetric" paraxial mode (Fig.\ \ref{F:Diagram2}), defined by the annihilation operator,
\begin{equation}\label{E:Symmetric}
	\hat{a}_+ = \frac{1}{\sqrt{N}} \sum_{i = 1}^{N} \hat{a}_i.
\end{equation}
This mode defines a laser pulse of duration $\tau$.   In calculating the tangle in this decomposition all of the other field modes, nonparaxial and nonsymmetric paraxial modes, act as a reservoir which, when traced over, lead to a mixed-state description of our bipartite system.   We assume $\kappa \tau \ll 1$, implying that the probability for the atom to emit a photon into one of the paraxial modes is very small.  In the Mollow picture (used throughout)  the Hilbert space of the quantum field may thus be truncated, allowing at most one excitation in the paraxial subsystem. Note that the atom is still allowed to spontaneously emit an arbitrary number of photons into the nonparaxial field modes.  Under these assumptions the atom and the symmetric field mode both behave as two-level systems, coupled together in an overall mixed state.  The tangle for such two-level systems can be calculated using Wootters' general formula for two qubits \cite{Wootters1998}. 

We begin by calculating the atom-symmetric mode density matrix from Eq.\ (\ref{E:Master}) for  $\kappa \tau\ll 1$.  In principle, one could solve for the state of the entire atomic-paraxial system, and then trace out the nonsymmetric modes.  However, even limiting all paraxial modes to at most one total excitation there are still $N+1$ possible field states, $N$ being the number of coarse-grained modes.  A more sensible approach is to keep only the ``dynamic symmetric mode" at any instant (as defined below), tracing out any non-symmetric modes along the way.  

Consider first a single time step $\Delta t = \Delta z/c$, with $\Delta z$ the coarse-graining length.  At the start of this interval the system Hilbert space has dimension $2\otimes2$.  The system consists of the two-level atom and a quantized symmetric mode of the field composed {\em only} of those coarse-grained modes which have already passed the atom, and restricted to one possible excitation.  Assuming $n<N$ modes have passed, the annihilation operator for this symmetric field mode will be
\begin{equation}\label{E:symn}
	\hat{a}_\text{+,n} = \frac{1}{\sqrt{n}} \sum_{i = 1}^n \hat{a}_i.
\end{equation}
After evolution over a time interval $\Delta t$, we choose a new subsystem division so that we again have a $2\otimes 2$ system.  The second subsystem will now correspond to the symmetric mode over the $n+1$ coarse-grained modes that have now passed the atom.  The procedure may then be repeated to build up the full evolution, and the tangle can be calculated at each step.

The procedure for evolving the state from time $t$ to $t+\Delta t$ is as follows.  Suppose that $n< N$ time intervals have passed.  The density operator $\hat \rho_n$ will be spanned by the basis states
\begin{equation}\label{E:startbasis}
	\ket{e}\ket{0}_{+,n} \quad \ket{g}\ket{0}_{+,n}\quad
	\ket{e}\ket{1}_{+,n} \quad \ket{g}\ket{1}_{+,n}.
\end{equation}
During the next interval a new coarse-grained mode impinges upon the atom.  This mode starts in the vacuum (in the Mollow picture) so the new state for this composite system is
\begin{equation}
	\hat \rho = \hat \rho_n \otimes \ket{0}_{n+1}\bra{0}.
\end{equation}
Through the next time interval this state evolves under a combination of the Jaynes-Cummings Hamiltonian which couples the atom to the quantized mode, a semi-classical coherent atom-laser interaction, and spontaneous emission into the nonparaxial modes.  We model this by a short time integration of the master equation, Eq.\ (\ref{E:Master}).  During this time the density matrix must be expanded to include the newly interacting mode.  The basis states include the previous set, Eq. (\ref{E:startbasis}), tensored with $\ket{0}_{n+1}$ for the newly added mode, plus the two new states
\begin{equation}\label{E:newbasis}
	\ket{e}\ket{0}_{+,n} \ket{1}_{n+1}\quad \ket{g}\ket{0}_{+,n}\ket{1}_{n+1}.
\end{equation}
States of the form $\ket{1}_{+,n}\ket{1}_{n+1}$ are higher order terms that are ignored under the assumption that $\kappa \tau \ll 1$.  This gives a total of six basis states for the density operator.   After performing the evolution over this time interval we trace out the nonsymmetric modes of the extended paraxial subsytem.  To do this, we perform a unitary transformation on the field operators to create a new tensor product decomposition.  Instead of the decomposition described in Eqs.\  (\ref{E:startbasis},\ref{E:newbasis}) consisting of the $n^{\text{th}}$ symmetric mode $\otimes$ the $(n+1)^{\text{th}}$ coarse-grained mode, we choose symmetric and antisymmetric modes described by the annihilation operators
\begin{equation}
	\hat a_{\pm,n+1} = \sqrt{\frac{n}{n+1}} \hat a_{+,n} 
		\pm \frac{1}{\sqrt{n}} \hat a_{n+1}.
\end{equation}
The transformation which achieves this is
\begin{equation}
	U = \begin{pmatrix} 1&0&0\\0&\sqrt{\frac{n}{n+1}}&\sqrt{\frac{1}{n+1}}\\
	 0& - \sqrt{\frac{1}{n+1}} & \sqrt{\frac{n}{n+1}}\end{pmatrix}.
\end{equation}
Finally $\hat \rho_{n+1}$ can be obtained by tracing out the antisymmetric field mode.  

The mixed state entanglement between the atom and the symmetric field mode can be calculated using Wootters formula for the tangle \cite{Wootters1998} directly from this density matrix.  The results will depend on several physical parameters: the area of the beam, the intensity of the laser $|\alpha|^2$, and the duration of the pulse.  Additionally, the numerical accuracy will depend upon the length of the time step used in the calculations, or equivalently the number of time steps $N$.  To maintain the paraxial approximation, we choose the area of the beam to be $\bar{A}=A/\sigma_\text{eff} = 1000$.  This corresponds roughly to the quadrupole transition $S_{1/2}$ to $D_{5/2}$ in $^{40}\text{Ca}^+$ considered by van Enk and Kimble \cite{vanEnk2001}, for which a laser focused to a spot size of 100 $\mu \text{m}^2$ corresponds to $\bar{A}\approx 1600$.  For simplicity we consider only $\pi/2$ pulses, such that $g \alpha \tau = \pi/2$.  With the relations $\kappa = \Gamma / \bar{A}$ and $g = \sqrt{\kappa / \Delta t} = \sqrt{\kappa N /\tau}$, assuming a fixed $N$, we have a direct relationship between $\alpha$ and $\tau$.  We choose as our free parameter the emission probability, $\Gamma \tau$, which sets the degree of entanglement between the atom and laser.

\begin{figure}
	\scalebox{.45}{\includegraphics{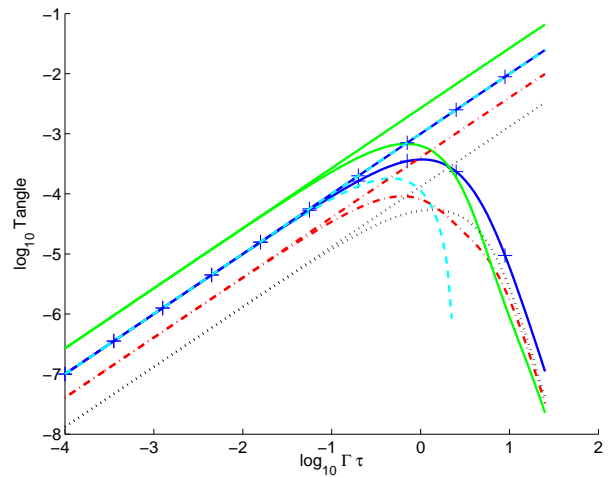}}
	\caption{The entanglement (tangle) generated between a two-level atom and a quantized $\pi/2$ laser-pulse versus the decoherence ($\Gamma \tau$) on a log-log plot. The pulse consists of $N = 10^4$ coarse-grained modes and a normalized beam area $\bar{A }= 10^3$ .  Including spontaneous emission out of the beam, the tangle grows as a simple power law for $\Gamma \tau \ll 1$, then drops exponentially fast for large decoherence.  Shown for comparison are the results of the tangle calculated using a closed system Jaynes-Cummings dynamic that grown linearly for all times on this plot, but agree very well with the full calculation when the probability of spontaneous emission is small.  Six different initial atomic conditions are shown: $\ket{e}$ (blue, solid cross), $\ket{g}$ (green, dashed), $\ket{e} + i \ket{g}$ (cyan, solid), $\ket{e} - i  \ket{g}$ (black, dotted), $\ket{e} \pm \ket{g}$ (red, dash-dot).  These last two initial condition show exactly the same tangle for all times (see text).\label{F:SymmetricTangle}}
\end{figure}

A plot of the tangle versus $\Gamma \tau$ is given in Fig. \ref{F:SymmetricTangle}.  Each point in the plot corresponds to the total tangle measured just after the $\pi/2$ pulse has completely passed the atom.  For very short pulses, the growth in the tangle is linear with $\Gamma \tau$ for all initial conditions, leading to a unit slope on the log-log plot.  However the constant of proportionality is different.  This difference can be attributed to the probability of the atom being in the excited state during the pulse interval -- the more excitation the higher the rate of growth.  With a greater probability of spontaneous emission, the behavior deviates strongly from this linear trend, peaking roughly where the pulse length and decoherence time are comparable.  After this time the entanglement drops precipitously, and for at least one initial condition, the atom and symmetric field mode become separable.  The maximum emission parameter shown in the plot is $\Gamma \tau = \sqrt{\bar{A}}=\sqrt{1000}$, which corresponds to $\kappa \tau = 1/\sqrt{\bar{A}}=1/\sqrt{1000}$; for a larger emission probability, our truncation of the field Hilbert space to one excitation breaks down.

\subsection{Approximations}
The calculation in Sec.\ \ref{S:SymCalc} yields the entanglement between the atom and the laser mode as it propagates past the atom.  It is a complex procedure because we have explicitly separated the two sources of decoherence: spontaneous emission into the nonparaxial modes and spontaneous emission into the paraxial but nonsymmetric modes.  We did this because spontaneous emission into the symmetric-paraxial mode is the source of the entanglement and we must be careful not neglect this important effect.  Nonetheless, emission into the nonsymmetric-paraxial modes is a small fraction of the total loss since $\kappa \ll \Gamma'$.  We can consistently maintain the coherent coupling between the atom and the symmetric mode while neglecting emission into the nonsymmetric mode since it acts only as a small source a decoherence, a process dominated by off-axis emission. We thus lump all sources of decoherence together, with total rate $\Gamma = \kappa + \Gamma'$.  In this case the Master equation becomes,
\begin{equation}\label{E:DecMaster}
	\totald[\hat{\rho}]{t} = \frac{-i}{\hbar} [\hat H',\hat{\rho}] + \frac{\Gamma}{2} \left(2 \hat{\sigma}_- \hat{\rho} \hat{\sigma}_+ - \hat{\sigma}_+ \hat{\sigma}_- \hat{\rho} - \hat{\rho} \hat{\sigma}_- \hat{\sigma}_+ \right),
\end{equation}
where $\hat H'  = -\hbar \Omega (\hat \sigma_{+} + \hat \sigma_{-} )/2  - \hbar g (\hat{a} \hat \sigma_{+} + \hat{a}^{\dag} \hat \sigma_{-} )$ is the driven Jaynes-Cummings Hamiltonian.  This equation describes a cavity QED dynamic plus decay into a reservoir \cite{Kimble1998}.   When the evolution of the tangle is calculated using this equation of motion,  the results are in complete agreement with those obtained in Sec.\ \ref{S:SymCalc} to within numerical accuracy.

A more radical simplification is to ignore the decoherence term in Eq.\ (\ref{E:DecMaster}) altogether when the interaction time is small compared to the atomic lifetime.  The remaining dynamics represents a closed system, with the atom and the symmetric mode evolving unitarily. 
\begin{equation}\label{E:SimpleMaster}
	\partiald[\hat{\rho}]{t} = \frac{-i}{\hbar} [\hat H',\hat{\rho}] 
\end{equation}
For this evolution, the initial bipartite pure state remains pure and the entanglement can be calculated in a straightforward manner.  Such an approximation is routinely made for atoms in cavities in the strong coupling regime, where $g \gg \Gamma$ \cite{Kimble1998}.  In free space we can express $g$ in terms of $\kappa$; the strong coupling approximation requires $\kappa \gg \Gamma^2 \tau$, or equivalently $\Gamma \tau \bar{A} \ll 1$.  The beam area may be chosen arbitrarily large, however, so that this inequality need not hold.  One is  \textit{not} in the strong coupling regime in free space and so agreement between the entanglement calculated using closed and open system models should not be expected apriori.  This has been the source of some recent controversy \cite{Itano2003,vanEnk2003,Banacloche2003}.

To compare the numerical results obtained using Eq.\ (\ref{E:Master}) with those obtained using the closed system evolution Eq.\ (\ref{E:SimpleMaster}) we first calculate the tangle analytically using the closed system dynamics.  Though one could measure the closed system entanglement using the von Neumann entropy of the marginal state as was done in \cite{vanEnk2001}, for direct comparison with our results, we use the tangle given in Eq.\ (\ref{E:PureTangle}).  A simple calculation yields,
\begin{equation}\label{E:TangleFormula}
	T = \kappa \tau \frac{\left(2\E{\sigma_y}+\pi\right)^2 +\E{\sigma_x}^2\left(4 - \pi^2\right)}{\pi^2} + O\left((\kappa \tau)^2\right), 
\end{equation}
where the Pauli matricies describe the two-level atom in the usual way.  Expectation values in this equation are to be taken using the initial state of the atom.   The results of this approximation are shown in Fig.\ \ref{F:SymmetricTangle}.  For the parameters where all approximations are valid, we see a simple linear growth of the entanglement with emission probability.  Though slightly larger, the entanglement calculated via the closed system dynamics is in very good agreement, $O(\Gamma \tau)$, with our numerical results for the full atom-symmetric field mode tangle when $\Gamma \tau \ll 1$, as seen in Fig.\ \ref{F:SymmetricTangle}.  This agreement holds even when the system is not in the strong coupling regime, $\bar{A} \Gamma \tau > 1$.

\begin{figure}
	\scalebox{.45}{\includegraphics{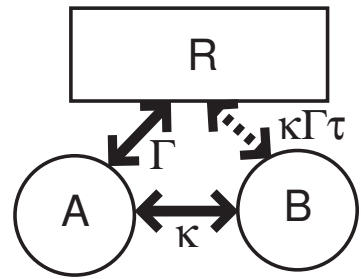}}
	\caption{The decoherence channels for the atom-laser mode interaction are shown, with each block representing a subsystem of the entire Hilbert space. Subsystem $A$ (the atom) is connected directly to reservoir $R$ (all vacuum modes) with coupling rate $\Gamma$ (the spontaneous emission rate).  System $B$ (the laser mode) is connected to the reservoir $R$ only indirectly through system $A$.  Since $B$ couples to $A$ at rate $\kappa$ (the measurement strength), it is indirectly coupled to $R$ at rate $\Gamma \kappa \tau$, where $\tau$ is the duration of the interaction. \label{F:Diagram}}
\end{figure}

To understand why the entanglement calculated using closed and open system dynamics agree so well even when the strong coupling assumption is violated, consider the processes which lead to the loss of entanglement.  The inherent dynamics of the atom-laser mode system can be represented by a bipartite system, with subsystems $A,B$, coupled to a reservoir of other electromagnetic modes, $R$,  as depicted in Fig.\ \ref{F:Diagram}.  System $A$ is coupled to system $B$ with strength $\kappa$, while it is coupled to the reservoir $R$ with strength $\Gamma$.  System $B$ is not coupled directly to the reservoir, coupling only indirectly through system $A$.  The total system consisting of both subsystems $A,B$ will start in some  pure state.   From the perspective of quantum trajectories \cite{Carmichael1993}, there are two possible ways in which the reservoir $R$ can reduce the entanglement generated by the dynamics between $A$ and $B$.  In the first method  the nonunitary but deterministic no-jump evolution of system $A$ leads to a reduction of this entanglement with probability $1-e^{-\Gamma \tau}$.  The total loss of entanglement is thus proportional to $\Gamma \tau T$ for $\Gamma \tau \ll1$, where $T$ is the tangle generated in the closed system dynamics between $A$ and $B$.  Since entanglement will be generated in the closed system at a rate proportional to the coupling $\kappa$, the total loss of entanglement is of order $O(\Gamma \tau \kappa \tau)$.  Alternatively, a nondeterministic jump in a quantum trajectory can reduce entanglement if it leads to a particular statistical mixture of states as described below.  The probability of this particular type of correlated quantum jump is, however, limited by the smaller decoherence rate of the two subsystems $A,B$.  In this case system $B$ can only decay through system $A$ in the interval $\tau$ at a rate proportional to $\Gamma \tau \kappa$.  The amount of entanglement that can be destroyed through this process will then be of order $O(\Gamma \tau \kappa \tau)$.  Thus both deterministic decay and quantum jumps lead to entanglement loss of order $\Gamma \tau \kappa \tau$.  This is sufficiently small to ensure that the closed system dynamics generates the same entanglement as the open system dynamics when $\Gamma \tau \ll 1$ and $\kappa \tau \ll 1$, independent of the relation between them, i.e. even when the strong coupling approximation breaks down.

The above explanation can be made rigorous for the atom-symmetric field mode system.  Assume that $\Gamma \tau \ll1$ and $\kappa \tau \ll 1$, though $\bar{A} \Gamma \tau$ may be large.  For $\Gamma \tau \ll 1$ the state of the system after the pulse has passed may be represented by a density operator of the form
\begin{equation}\label{E:rhosplit}
	\hat{\rho} = (1-\epsilon) \ket{\psi}\bra{\psi}+ \epsilon \hat \rho_{\text{jump}}.
\end{equation}
Here $\ket{\psi}$ is the state evolved under the closed system unitary dynamics, while $\hat \rho_{\text{jump}}$ represents the state conditioned on a jump having occurred during the interaction time.  The probability for a jump is denoted $\epsilon$ and will be proportional to $\Gamma \tau$.  These assumptions alone are insufficient to ensure that entanglement is preserved.  As an example, consider the entangled state $\ket{\psi} = (1-\xi/2)\ket{g 1} + \sqrt{\xi} \ket{e 0}$.  This state has tangle $T \approx 4 \xi$ for $\xi \ll 1$.  Note that the state $\ket{\psi}$ is equivalent under local unitaries to any other state with the same entanglement.  If we choose $\hat{\rho}_\text{jump} = 1/2 \ket{e 1}\bra{e 1} + 1/2\ket{g 0}\bra{g 0}$, the statistical mixture $\hat \rho$ given in Eq.\ (\ref{E:rhosplit}) is separable for $\sqrt{\xi}\leq \epsilon$, as can be seen by its positive partial transpose \cite{Peres1996,Horodecki1996}.  For our particular case, an atom in free space, $\sqrt{\xi} \leq g \tau$, since entanglement is generated by the atom-field coupling.  So for this choice of $\hat{\rho}_{\text{jump}}$, the system can be made separable when $g < \Gamma$, or equivalently $\bar{A} \Gamma \tau >1$.  

Since our calculations shows that the open system maintains the same entanglement as the closed system, even when $\bar{A} \Gamma \tau>1$, the dynamics must be restricted  such that these special statistical mixtures are rare.  To see this, note that the post-jump state we considered above must be generated by a highly correlated jump process.  The probability of such a process is very low.  Indeed, to good approximation, we expect $\hat{\rho}_\text{jump}$ to be of the form
\begin{equation}\label{E:RhoDec}
	\hat \rho_\text{jump} = \hat \rho_{\text{atom}} \otimes \ket{\text{vac}}\bra{\text{vac}}
\end{equation}
where $\hat \rho_{\text{atom}}$ is some arbitrary (possibly mixed) state of the atomic subsystem.  This ignores the process of spontaneous emission into a nonparaxial mode followed by subsequent excitation of a paraxial mode.  This higher order process may be safely ignored when we are considering entanglement of order $\kappa \tau$, and  $\sqrt{\kappa \tau} \gg \kappa \tau \Gamma \tau$, which is satisfied when $\Gamma \tau \ll 1$.  Inserting Eq.\ (\ref{E:RhoDec}) into Eq.\ (\ref{E:rhosplit}) the tangle may then be calculated using Wootters' formula, assuming that $\ket{\psi}$ is an arbitrary state that can be generated using the closed system dynamic.  To lowest order, $O(\kappa \tau \Gamma \tau)$, this agrees with the tangle obtained using closed system dynamics.

\subsection{Bounding the Total Entanglement \label{S:FullTangle}}
We now consider the entanglement that is generated between the atom and all paraxial modes.  This is the physically relevant quantity that determines the error rate in atom-laser control, modulo spontaneous emission into the nonparaxial modes.  In addition, it gives a measure of the dynamically generated entanglement resource that is accessible in principle.  A calculation of this total entanglement is significantly more complex than the calculation performed in Sec.\ \ref{S:Symmetric} as the Hilbert space associated with all coarse-grained paraxial modes is much larger than that of the single symmetric mode. Even if we assume $\kappa \tau \ll 1$, and thereby restrict paraxial field modes to a single excitation as in Sec.\ \ref{S:Symmetric}, the remaining Hilbert space has dimension $2\otimes(N+1)$.  A calculable measure of entanglement for a mixed state of a $2 \otimes N$ system for $N>2$ is not known except for some special cases which do not apply here \cite{Rungta2003,Vollbrecht2001}.  We thus resort to calculating a bound for the total entanglement, using the tangle as our measure.

The general difficulty with calculating the entanglement of a mixed state arises because the ensemble decomposition of a density operator into a statistical mixtures of pure states $\hat{\rho} = \sum_ i p_i \ket{\psi_i}\bra{\psi_i}$ is not unique.  Averages of a pure state entanglement monotone for different ensemble decompositions do not usually lead to the same result and one must thus perform a difficult minimization search to find the actual entanglement.  Since the tangle is a convex function, however, any ensemble decomposition provides an upper bound according to,
\begin{equation}\label{E:Convex}
	T(\hat\rho) 
		\leq \sum_i p_i T\left(\ket{\psi_i}\bra{\psi_i}\right) \forall \{\psi_i,p_i\}
\end{equation}
Given a known ensemble decomposition, we can calculate the right hand side by averaging the tangle for the given bipartite pure states, over the appropriate probability distribution, using Eq. (\ref{E:PureTangle}).

A natural ensemble decomposition is obtained by integrating the master equation of the atom-paraxial system using the quantum trajectory method \cite{Carmichael1993}. A stochastic Schr\"{o}dinger equation describes the evolution of an open quantum system with nonunitary evolution due to its coupling to the environment, and conditioned on a fictitious measurement performed on the environment with unit efficiency.  For each measurement record $M(t)$ there is a corresponding pure state trajectory $\ket{\psi_M(t)}$.  Averaging over the all possible measurement records results in a mixed state description, with density operator given by the convex combination
\begin{equation}
	\hat \rho (t)= \sum_M p_{M}(t) \ket{\psi_M(t)}\bra{\psi_M(t)}.
\end{equation}
From this construction one may immediately write down an upper bound on the tangle using Eq. (\ref{E:Convex}).

Brute force calculation of all the required trajectories is straight forward but prohibitive.  If we choose the fictitious measurement to be photodetection, a measurement record will consist a sequence of bits denoting count/no-count decisions in each interval $(t,t+\Delta t)$ (ignoring the possibility of two or more photodetections as it is of order $\delta t^2$).  Assuming there are $N = 1000$ such intervals then there are $2^{1000}$ trajectories.  Our calculation is made tractable, however, by noting that only the most recent jump in any trajectory will contribute to the tangle; the state of any field modes which passed the atom prior to the last jump will not be entangled with the atom since the post-jump state is separable (the atom in the ground state and the field in a coherent state).  The evolution after a jump has occurred will be independent of when the jump occurred as the system is Markov, and all of the quantized paraxial modes prior to interaction with the atom are in the same coherent state. It is thus sufficient to calculate two no-jump trajectories -- the initial trajectory $\ket{\psi_{A}(i)}$ and the post-jump trajectory $\ket{\psi_{B}(i)}$.  Both represent deterministic evolution according to a nonHermitian Hamiltonian, including the coupling of the system to the nonparaxial modes assuming \textit{no quantum jumps} into the nonparaxial modes.  They differ only in their initial conditions: $\ket{\psi_{A}(i)}$ is the evolution given the initial state of the atom at $t=0$ and $\ket{\psi_{B}(i)}$ is the evolution starting with the atom in the ground state right after a jump.  An upper bound on the tangle then follows from Eq. (\ref{E:Convex}),
\begin{equation}\label{E:JumpTangle}
	T(\hat \rho) \leq  \sum_{1=1}^{N} P_{LJ}(i) T_{PJ}(N-i) + P_{NJ} T_{NJ},
\end{equation}
where $P_{LJ}(i)$ and $P_{NJ}$ are respectively the probabilitiy that the last jump occurred during the time interval $i$ and the probability that no jump occurred during the entire evolution.   The tangle associated with no jumps during any interval follows from Eq.\ (\ref{E:PureTangle}), $ T_{NJ} = T\left(\ket{\psi_A(N)}\right)$, whereas the post-jump tangle is given by $T_{PJ} (N-i) = T\left(\ket{\psi_B(N-i)}\right)$.

The relevant probabilities that appear in Eq.\ (\ref{E:JumpTangle}) may also be calculated using the fiducial trajectories.  In particular under nonunitary evolution without normalization, the probability that no jump occurs in a given trajectory up to time $t$ is $\| \tilde{\psi} (t) \| ^2$.  Thus, the probability that no jump occurs during any of the $N$ intervals is $P_{NJ}= \| \tilde{\psi}_A (N) \|^2 $.  The probability that the last jump occurred in interval $i$ is given by,
\begin{equation}
	P_\text{LJ}(i) = P_{NJ}(N|i)  P_J(i),
\end{equation}
where $P_{NJ}(N|i) = \| \tilde{\psi}_B (N-i) \|^2 $ is the conditional probability that no jumps occur between the interval $i$ and $N$ given that a jump occurred in interval $i$. The total probability that a jump occurred at time step $i$, $P_J(i)$, will satisfy the following equations:
\begin{subequations}
\begin{align}
	P_J (i) &= P_A(i) + \sum_j  P_J (i|j) P_A(j) ,\label{E:FirstJump}\\
	P_J (i|j) &= P_B(i-j) + \sum_{k>j}  P_J (i|k)P_B(k-j).\label{E:SecJump}
\end{align}
\end{subequations}
The first equation expresses the total probability for a jump in interval $i$ as the sum of the probability that the system had its first jump in interval $i$, $P_A(i) $, plus the probability that it first jumped in interval $j<i$ then some time later jumped in interval $i$.  $P_{J}(i|j)$ is thus the conditional probability for a jump at $i$ given a previous jump occurred in interval $j$.  Equation (\ref{E:SecJump}) provides a recursive relation for $P_{J}(i|j)$ in terms of, $P_B(i-j)$, the probability for a jump to occur $i-j$ intervals after a previous jump occurred, with no intervening jumps.  The probability $P_A(i)$ can be easily calculated from the fiducial trajectory $\ket{\psi_A(i)}$ and similarly $P_B(i-j)$ can be calculated from $\ket{\psi_B(i-j)}$ \cite{Carmichael1993}.  Since $P_J  (i|j)$ is only a function of $i-j$, as follows from the Markov property, the recursion relation (\ref{E:SecJump}) can be solved independently. We find a suitable solution by truncating the recursion at $n$ steps corresponding to at most $n+1$ jumps into the nonparaxial modes.  Solving for $P_{LJ}(i)$, using this truncated solution, we can then obtain an upper bound on the tangle according to Eq.\ (\ref{E:JumpTangle}).

\begin{figure}
	\scalebox{.45}{\includegraphics{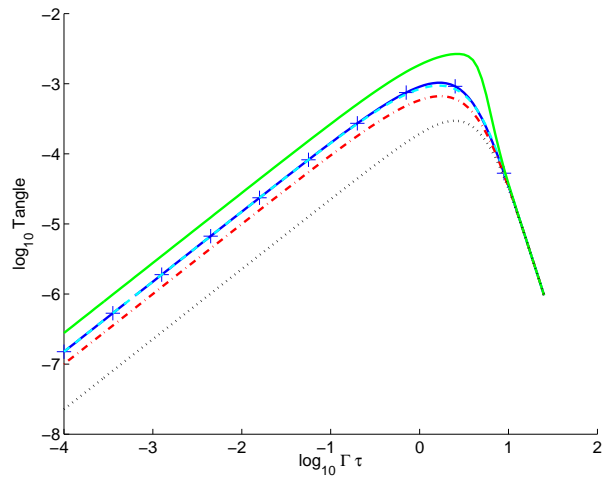}}
	\caption{An upper bound on the the entanglement (tangle) generated between the atom and all coarse-grained paraxial field modes is plotted versus the total decoherence ($\Gamma \tau$) in a log-log plot.  The different curves represent different atomic intial conditions with parameters given in Fig. \ref{F:SymmetricTangle}   \label{F:FullTangle}}
\end{figure}

Our upper bound on the tangle is plotted in Fig.\ \ref{F:FullTangle} as a function of time.  For short times, linear behavior is exhibited, as in the previous calculations for the symmetric tangle.  Note, however, that the rate of entanglement generation is always greater than the symmetric calculation.  For short pulse durations $\Gamma \tau \leq 10^{-2}$ there is only a $5\%$ difference in rates for the initial condition $\ket{g}+i\ket{e}$,  the state that is most excited during the interval.  All of the plotted upper bounds turn over at roughly the same point, slightly past where the pulse duration equals the decoherence time, and then converge on a single limit for large decoherence probability.  This behavior is an artifact our calculation --  for long times all information about the initial state has been lost since all trajectories with appreciable probability have undergone multiple jumps.   

\section{Summary}
Entanglement is generated between a laser beam and an atom in free space due to the atom's spontaneous emission of photons into the paraxial modes, superimposed on the laser's coherent field amplitude. Through the use of judicius approximations, we  were able to quantify this entanglement, a fundamentally hard problem for a large dimensional open quantum system. In particular we calculated the entanglement between the atom and the field mode defined by the laser pulse. Of particular interest is how our measure of a entanglement is reproduced under further simplifying assumptions.  We considered two models: a lumped decoherence model and a closed-system model.  In the former all sources of decoherence, paraxial and nonparaxial, are lumped into a single decoherence term.  This is in excellent agreement with the full calculation when the paraxial approximation holds, $\bar{A}\gg1$.  This is as expected, since the vast majority of the decoherence is due to the nonparaxial modes.  More surprisingly, a closed system model which treats the quantized atom laser-pulse dynamics via a single mode Jaynes-Cummings Hamiltonian is in excellent agreement when $\Gamma \tau \ll 1$, even when the strong coupling assumption was violated.  This result was unexpected since the naive picture would argue that, in the absence of the strong atom-mode coupling, $\bar{A} \Gamma \tau >1$, spontaneous emission into other modes should destroy entanglement.  We have shown, however, that the particular form of the coupling between the vacuum modes and the atom-laser system in fact \textit{preserves} the entanglement.  Thus, there is no contradiction between the observation that all quantum effects in the atom-laser interaction are due to spontaneous emission \cite{Itano2003} and the fact that atom-laser entanglement can be accurately modeled using a single mode Jaynes-Cummings dynamic \cite{vanEnk2001,Banacloche2002} when $\Gamma \tau \ll 1$.  Such agreement is ensured by the fact that spontaneous emission is the \textit{only} decoherence mechanism. 

Finally, we placed an upper bound on the total entanglement that is generated between the atom and the full set of paraxial field modes.  We used the quantum trajectory method to calculate a dynamical unravelling of the master equation, which provides a natural ensemble decomposition of the bipartite density matrix.  Averaging the pure state tangle over this ensemble leads to an easily calculated upper bound. It remains to be seen how tight this bound is.  Nonetheless, it provides a useful benchmark on the total entanglement resource that is available for quantum information processing applications.

\begin{acknowledgments}
The authors would like to thank Tracey Tessier and Aldo Delgado-Hilgado for useful discussions and suggestions.  This work was supported by NSF Grant No.~PHY-009569 and the National Security Agency (NSA) and the Advanced Research and Development Activity (ARDA) under Army Research Office (ARO) Contract No.~DAAD19-01-1-0648.  AS was supported in part as a QuaCGR Fellow under ARO Contract No.~DAAD119-02-1-0213.
\end{acknowledgments}

\end{document}